\documentclass{iau}
\usepackage{graphicx, aas_macros, url}
\usepackage[numbers]{natbib}

\title[Modeling with the Crowd] 
{Modeling with the Crowd: \\Optimizing the Human-Machine Partnership with Zooniverse}

\author[Hugh Dickinson, Lucy Fortson, Claudia Scarlata, Melanie Beck, Mike Walmsley]   
{Hugh Dickinson$^1$, Lucy Fortson$^1$, Claudia Scarlata$^1$, Melanie Beck$^1$ \& Mike Walmsley$^2$}

\affiliation{$^1$School of Physics and Astronomy, University of Minnesota, 116 Church Street SE, Minneapolis, MN 55455, USA
 \\[\affilskip]$^2$Oxford Astrophysics, Denys Wilkinson Building, Keble Road, Oxford OX1 3RH, UK
}

\pubyear{2019}
\volume{341}  
\jname{PanModel2018 : Challenges in Panchromatic Galaxy Modelling with Next Generation Facilities}
\editors{M. Boquien, E. Lusso, C. Gruppioni \& P. Tissera}
\begin{document}

\maketitle
\vspace{-5pt}
\begin{abstract}
 LSST and Euclid must address the daunting challenge of analyzing the unprecedented volumes of imaging and spectroscopic data that these next-generation instruments will generate. A promising approach to overcoming this challenge involves rapid, automatic image processing using appropriately trained Deep Learning (DL) algorithms. However, reliable application of DL requires large, accurately labeled samples of training data. Galaxy Zoo Express (GZX) is a recent experiment that simulated using Bayesian inference to dynamically aggregate binary responses provided by citizen scientists via the Zooniverse crowd-sourcing platform in real time. The GZX approach enables collaboration between human and machine classifiers and provides rapidly generated, reliably labeled datasets, thereby enabling online training of accurate machine classifiers. We present selected results from GZX and show how the Bayesian aggregation engine it uses can be extended to efficiently provide object-localization and bounding-box annotations of two-dimensional data with quantified reliability. DL algorithms that are trained using these annotations will facilitate numerous panchromatic data modeling tasks including morphological classification and substructure detection in direct imaging, as well as decontamination and emission line identification for slitless spectroscopy. Effectively combining the speed of modern computational analyses with the human capacity to extrapolate from few examples will be critical if the potential of forthcoming large-scale surveys is to be realized.
\end{abstract}
\vspace{-10pt}
\firstsection 
\section{Astronomy in the age of Big Data}
The coming decade promises an unprecedented expansion in the volume and diversity of high-quality data that will be available to the astronomical community. Forthcoming instruments including the Large Size Synoptic Telescope (LSST) \citep{ivezic2008lsst} or \textit{Euclid} \citep{2011arXiv1110.3193L} will provide galaxy catalogues containing billions of sources. The potential scientific yield of these datasets is correspondingly unprecedented as long as they can be rapidly and effectively analysed.

To be robust, traditional automated analyses of extensive astrophysical datasets have remained somewhat rudimentary. For example, in the context of galaxy morphology, commonly applied techniques for automatic classification use the radial light profile of galaxies to estimate the relative contributions from distinct bulge and disk components \citep[e.g.][]{2011ApJS..196...11S}. While such methods have largely solved the problem of \textit{coarse} morphological classification for large populations of galaxies, they cannot identify more subtle features like stellar shells, spiral arms, bars and clumps. Indeed, automatic detection of galaxy substructure is notoriously difficult, which often limits related studies to small, visually inspected samples \citep[e.g.][]{2010ApJS..186..427N}.

To visually inspect and morphologically categorize large numbers ($\sim10^5$) of galaxy images, the \textit{Galaxy Zoo} project \citep[hereafter GZ;][]{2008MNRAS.389.1179L} used a web-based classification interface to recruit over 350,000 volunteers from the public. Unfortunately, even the impressive resources mobilized by GZ will not be sufficient to process the enormous volumes of data that forthcoming instruments will deliver. To meet this challenge, robust new automatic tools that provide at least human-equivalent performance when applied to complex astrophysical analyses are urgently required.

%
Modern machine learning algorithms are designed specifically to process and analyse large data volumes associated with Big Data. Given sufficient training data, DL techniques can achieve human-like performance across a wide variety of image analysis tasks including morphological classification of galaxies \citep[e.g.][]{2015MNRAS.450.1441D,2018MNRAS.476.3661D}. The main obstacle for applying DL in any context is the requirement for abundant labeled training data. In this proceeding, we outline methods that leverage citizen science to generate accurate data labels in sufficient volume to train robust DL algorithms. We use examples based on GZ and use results of Galaxy Zoo Express \citep[hereafter GZX;][]{2018MNRAS.476.5516B} and planned extensions thereof to illustrate the techniques we discuss. We address three crucial questions:
\begin{itemize}
 \item How can the compromise between label accuracy and volunteer effort be optimized?
 \item How can we avoid volunteers labelling mundane data that are well characterized by DL models?
 \item How can the unique capabilities of humans and DL best be combined?
\end{itemize}

\vspace{-15pt}
\section{Optimizing Accuracy in Citizen Science}
\begin{figure}[!t]
 \begin{center}
  \includegraphics[height=1.7in]{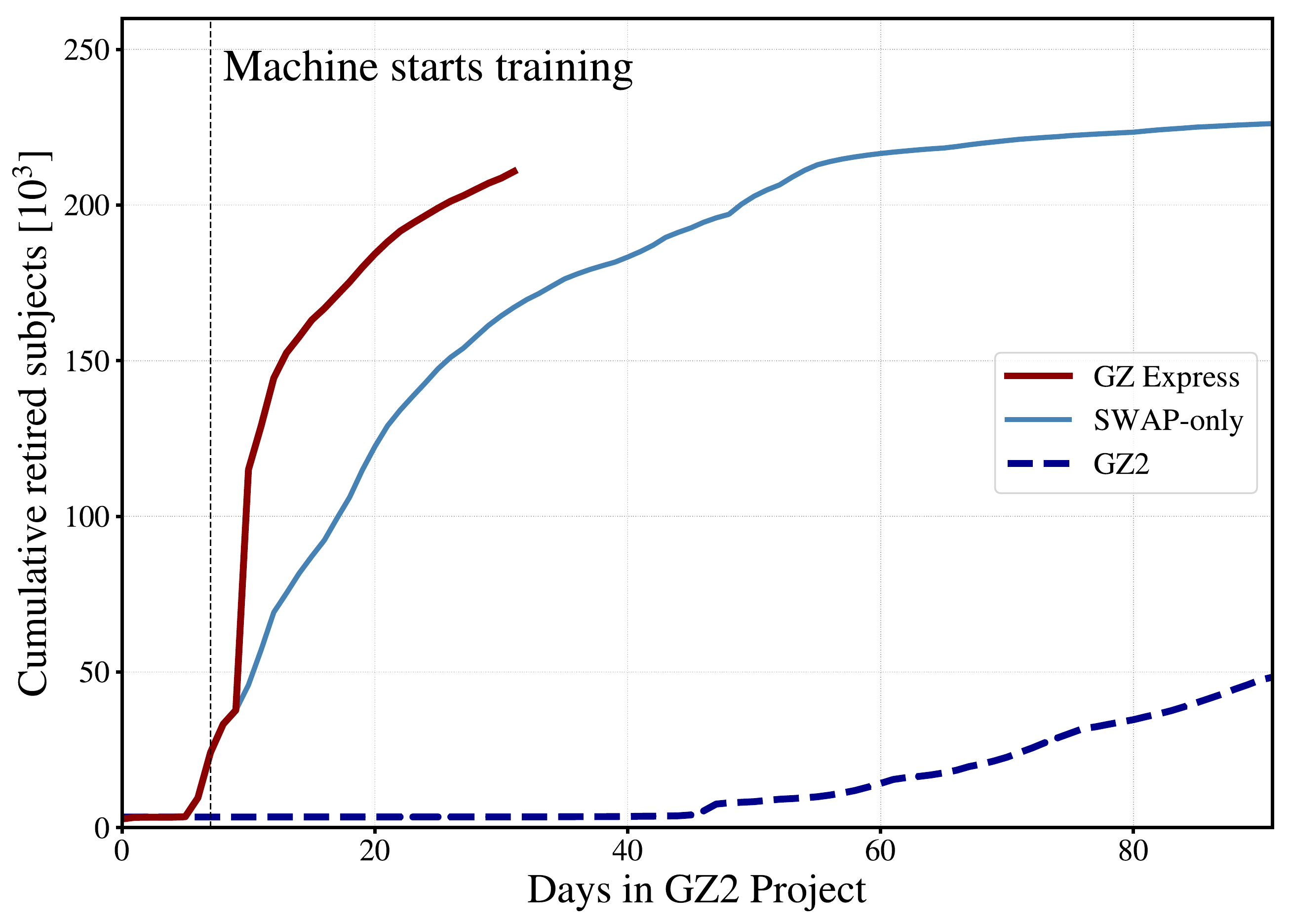}
  \includegraphics[height=1.7in]{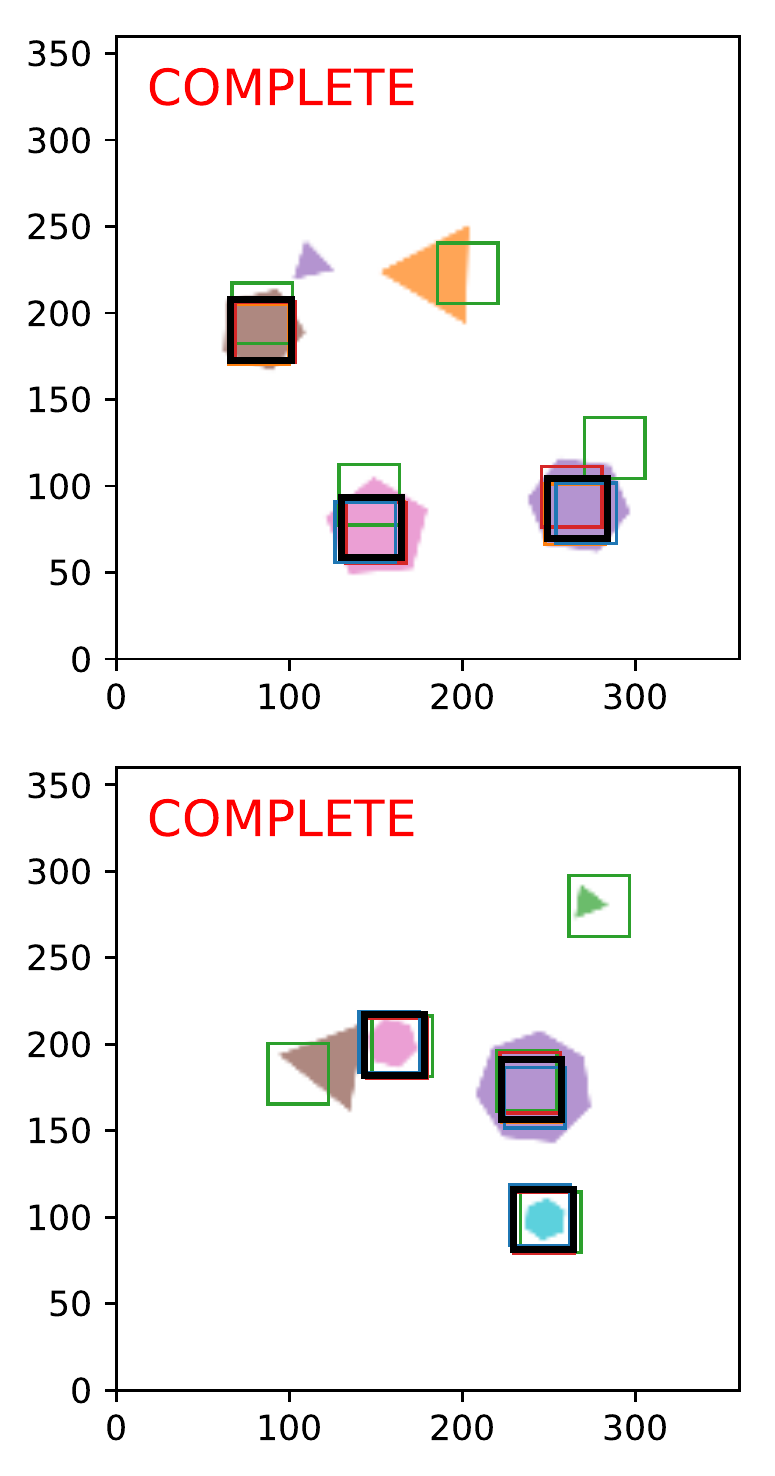}
  \includegraphics[height=1.7in]{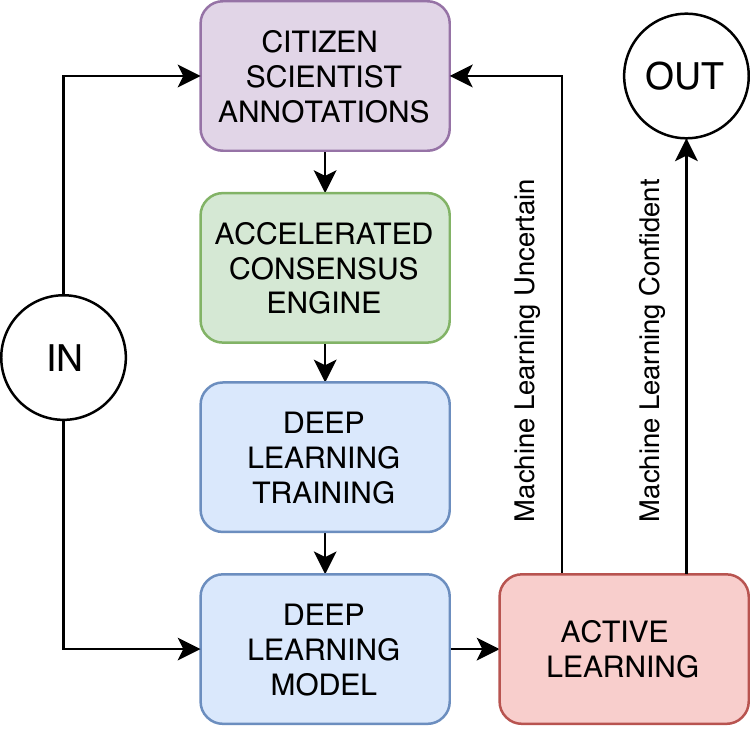}
  \caption{Optimizing the utility of citizen scientists' annotations using. \textit{Left:} Adapted from \cite{2018MNRAS.476.5516B}. The \textit{dashed} curve illustrates the cumulative retirement rate in the original GZ2 project. GZ2 required 40 annotations in order to retire each subject. The \textit{solid blue} curve illustrates the same quantity using the SWAP algorithm with a retirement threshold $P_{\mathrm{correct}}$ that replicates $>95\%$ of the original GZ2 labels. The \textit{solid red} curve illustrates the cumulative retirement rate using a simple random forest algorithm trained using a subset of the consensus classifications provided by SWAP within the first 12 days of volunteer activity. \textit{Middle:} Results demonstrating the effectiveness of an optimal aggregation algorithm for two-dimensional annotations. Volunteers are instructed to mark shapes with more than four sides. \textit{Thick black} boxes illustrate the consensus after annotation by four volunteers. The consensus annotations for both subjects are accurate. The coloured boxes indicate the annotations provided by individual volunteers.
   Evidently, the consensus algorithm is robust against the noisy or spurious \textit{green}-coloured annotations. \textit{Right:} Data flow schematic illustrating an infrastructure designed to enable efficient human-machine collaboration for the analysis of extensive astronomical datasets. The schematic unites a GZX-like infrastructure involving citizen science (\textit{purple}), optimal subject retirement (\textit{green}) and machine learning (\textit{blue}) components with an active learning engine (\textit{red}) designed to optimise DL model training.}
  \label{fig:gzx}
 \end{center}
 \vspace{-5pt}
\end{figure}

Aggregation of annotations that are provided by different citizen scientists \textit{for the same subject} is essential to mitigate individual subjectivity and compensate for outliers and catastrophic misclassifications \citep[see e.g.][for some GZ-related examples]{2008MNRAS.389.1179L,2013MNRAS.435.2835W,2016MNRAS.461.3663H}. While aggregation can be performed \textit{a posteriori} using a fixed number of independent classifications per subject, this approach is generally inefficient because stable consensus between classifiers is often achieved before the prescribed number of annotations have been gathered.
The \textit{SpaceWarps} project \citep{2016MNRAS.455.1171M} developed the \textit{SpaceWarps Analysis Program} (SWAP) for binary-valued classification tasks. SWAP dynamically computes Bayesian estimates for the probability $P_{\mathrm{correct}}$ that the current consensus label is correct, based on calibrated reliabilites of the volunteers that provided annotations. SWAP provides a quantitative means to optimize the number of annotations collected per subject by accepting the consensus once $P_{\mathrm{correct}}$ exceeds an appropriately chosen threshold. When the consensus is accepted the subject is said to be ``retired''. GZX applied SWAP to the morphological classification of galaxies by post-processing binary-valued volunteer responses from Galaxy Zoo 2 \citep[hereafter GZ2;][]{2013MNRAS.435.2835W} to the question ``Is the galaxy simply smooth and rounded, with no sign of a disk?''
Comparing the \textit{dashed blue} and \textit{solid blue} curves in the left panel of Figure \ref{fig:gzx} reveals that application of the SWAP algorithm increased the subject retirement rate by a factor of $\sim5$. The increased rate reflects the fact that most galaxies achieve stable consensus using far fewer than the default 40 annotations required by GZ. Indeed, application of SWAP reduced the overall volunteer effort by a factor of 7 while replicating $>95\%$ of the original GZ2 labels for the smooth versus disk classification.
The SWAP algorithm is designed to accelerate consensus for binary classifications, which restricts its domain of applicability to a subset of astrophysical use cases. In the context of galaxy morphology, many analyses require identification and labeling of specific galactic substructures, corresponding to specific locations or regions within a two-dimensional image. Optimal retirement algorithms for two-dimensional annotations have been developed in the context of commercial crowdsourcing \citep[e.g.][]{8100130} and we are currently adapting them to the Zooniverse (\url{www.zooniverse.org}) infrastructure that hosts GZ. The right panel in Figure \ref{fig:gzx} illustrates the effectiveness of one such algorithm using a simple test project in which volunteers were asked to mark all shapes with more than four sides. Reliable consensus was achieved after only four annotations for $>96\%$ of the test subjects. The results also demonstrate the algorithm's robustness against outliers and catastrophic mistakes.
\vspace{-15pt}
\section{Minimizing wasted volunteer effort}
Volunteer effort in citizen science is a valuable resource. It is not rational or ethical to waste volunteer time annotating subjects that DL algorithms can accurately characterize. A more sensible approach would focus volunteer attention on data that machine learning cannot handle, and use the resultant annotations as training data to refine the DL model.
\textit{Active Learning} (AL) is a general machine learning technique that incrementally updates a small initial training set using an \textit{acquisition function} (AF) to request annotations for subjects that, if annotated, would be most informative for the DL model in its current state. Subjects about which the model is currently uncertain are a natural choice. However, many conventional DL algorithms are incapable of quantifying the uncertainty associated with their model. Bayesian Convolutional Neural Networks \citep[BCNNs; e.g.][]{2015arXiv150602158G} are DL algorithms that are tailored for image analysis and which do provide an estimate of the model uncertainty required by AL in the form of approximate posterior distributions over the BCNN parameters.

Various AFs can be chosen to convert the model parameter distributions into subject uncertainty, each with different effects on how the model learns. A potential AF is the predictive entropy:
\vspace{-5pt}
\begin{equation}
 \mathbb{H}[y|\mathbf{x}, \mathbf{D}_{train}] =-\sum\limits_{c}^{}{p(y=c |\mathbf{x}, \mathbf{D}_{train})\log{p(y=c |\mathbf{x}, \mathbf{D}_{train})}}
 \vspace{-5pt}
\end{equation}
where $p(y=c |\mathbf{x}, \mathbf{D}_{train})$ is the probability that a trained DL model assigns subject $\mathbf{x}$ to class $c$ given the training dataset $\mathbf{D}_{train}$. Initial experiments using $\mathbb{H}$ as an AF have shown that large values of this metric generally correspond with broad BCNN posterior distributions, indicative of an uncertain DL model, while small $\mathbb{H}$ tends to indicate more concentrated BCNN posteriors. Moreover, $\mathbb{H}$ is also correlated with the accuracy of the BCNN model, with smaller values indicating a correct DL prediction is more likely.

\vspace{-15pt}
\section{Combined Analysis by Humans and Machines}
GZX explored the potential of human-generated classifications as training data for machine learning algorithms by using SWAP-accelerated consensus labels to train a simple random forest (RF) classifier. The \textit{red} line in Figure \ref{fig:gzx} reveals that the adequate initial training of the RF was possible within 8 days of volunteer activity. After 12 days, the SWAP provided over 40,000 labelled training images and the RF model was deemed sufficiently accurate to retire additional subjects based on its classifications. Using the RF classifier, over 200,000 subjects were retired within 27 days, increasing the overall classification rate by a factor of 10  while requiring $\sim13\times10^{6}$ fewer volunteer anootations with respect to GZ2. The RF classifications replicated $>93\%$ of the original GZ2 labels.

The right panel of Figure \ref{fig:gzx} illustrates a data flow schematic that unites and generalizes the infrastructure for accelerated consensus and machine learning training implemented in GZX. The schematic also includes an AL engine designed to optimize the DL training efficiency. We envision a system in which a small fraction of data are labeled by professional astronomers and citizen scientists. These labeled data are then used to train DL algorithms to perform specific analysis tasks. Partially trained DL algorithms classify unlabeled data and an active learning engine is used to estimate model uncertainty and request additional human annotations for poorly classified subjects. This design benefits from humans' unique ability to extrapolate reliably from very few examples and enables continual refinement of machine learning models as new and potentially unanticipated data classes are encountered. We reiterate that sophisticated automatic analyses will become essential tools for astronomy within the next few years and assert that infrastructures similar to that we have described will be invaluable to train, calibrate and validate such analyses.
\vspace{-15pt}
\bibliographystyle{iau_custom}
\bibliography{symposium_paper_revised}

\begin{discussion}
 \discuss{Katarzyna Ma\l{}ek}{What should we do when SWAP might not be able to classify more than a few classes of galaxies/objects? Should we then use unsupervised machine learning?}
 \discuss{Hugh Dickinson}{SWAP is designed to provide a quantitative estimate of the reliability of the consensus label after several citizen scientists have provided an annotation. If the probability of the consensus label indicates uncertainty (probability $\sim0.5$) after many volunteer classifications, then it may be that the particular image is particularly ambiguous. If this is true for a large fraction of your dataset, you may want to check that professional astronomers can provide a consistent classification for a representative sample of your data. If so, then unsupervised learning is an option, as is changing the question asked of citizen scientists - they may be struggling because they don't understand what is being asked of them.}
 \discuss{Dan Taranu}{In large datasets with millions of ``strange''/unclassifiable objects, how do you choose which ones should be sent back to astronomers for a second look? This was basically answered in the previous question as ``by finding clusters of outlier'', so follow-up: what if there are many small clusters of tens of rare objects?}
 \discuss{Hugh Dickinson}{There are several approaches being explored. Perhaps the most promising approach involved iterative refinement of initial clusters by human volunteers. By learning an embedding of the data space containing the poorly classified data clusters, nearby clusters can be presented to volunteers, whose classifications can be used to split, unify or purify each cluster. After several iterations we hope that very small clusters will merge enabling professional astronomers to investigate the larger cluster's unique characteristics.}
\end{discussion}

\end{document}